\documentclass[12pt]{article}
\usepackage{amsmath,amsfonts}
\usepackage{amscd,amsthm}
\usepackage{geometry}
\usepackage{graphicx}
\usepackage{color}
\usepackage[applemac]{inputenc}
\usepackage{tikz}
\usepackage{graphicx}
\usepackage{color}
\usepackage[applemac]{inputenc}
\usepackage{cases}
\usepackage{tikz}
\usepackage{algorithm2e}
\usepackage{setspace}
\usepackage{float}

\begin{document}

\title{Degree Variance and Emotional Strategies Catalyze Cooperation in Dynamic Signed Networks}

\author{Simone Righi \thanks{MTA TK "Lend\"{u}let" Research Center for Educational and Network Studies (RECENS),
Hungarian Academy of Sciences. Mailing address:
Orsz\'{a}gh\'{a}z utca 30, 
1014 Budapest, Hungary.
Email: simone.righi@tk.mta.hu}\and K\'{a}roly Tak\'{a}cs\thanks{MTA TK "Lend\"{u}let" Research Center for Educational and Network Studies (RECENS),
Hungarian Academy of Sciences. Mailing address:
Orsz\'{a}gh\'{a}z utca 30, 
1014 Budapest, Hungary.
Email: takacs.karoly@tk.mta.hu}}
%\date
\maketitle

\begin{abstract}
We study the problem of the emergence of cooperation in dynamic signed networks where agent strategies coevolve with relational signs and network topology. Running simulations based on an agent-based model, we compare results obtained in a regular lattice initialization with those obtained on a comparable random network initialization. We show that the increased degree heterogeneity at the outset enlarges the parametric conditions in which cooperation survives in the long run. Furthermore, we show how the presence of sign-dependent emotional strategies catalyze the evolution of cooperation with both network topology initializations.
\vspace{1cm}


\end{abstract}

\section{Introduction and Related Literature}

Cooperation among individuals is a key element for the survival and functioning of human and many animal societies. While cooperation is socially optimal, it is difficult to explain its existence in a population of selfish individuals. The Prisoner's Dilemma (PD) is frequently used to study this puzzle as it describes the situation in which the self interest of the individual is opposed to the emergence of cooperation. Two players are given two alternative strategies: to cooperate or to defect. Defection guarantees a higher payoff regardless of what the partner does and is thus the dominant strategy. However, cooperation - if played mutually - provides higher payoffs than mutual defection. 

A natural framework in which to study the emergence of cooperative behaviour is evolutionary game theory. This literature burgeoned following the seminar papers of \cite{maynardsmith1973the,maynardsmith1982evolution} and \cite{axelrod1981evolution}, with a large number of contributions being dedicated to the puzzle of cooperation (see \cite{rowthorn2009theories} for a recent survey of this subject). One strand of this literature looks into the effects of the structure of interactions on the outcome of the evolutionary process. In the context of the single shot PD, they find that cooperation in an unstructured population of randomly interacting individuals is not viable. Natural selection favors selfish defection, thus leading to groups composed entirely agents playing this strategy (\cite{taylor1978evolutionary,hofbauer1998evolutionary}). Introducing a more stringent structure for the social contacts and thus allowing only agents that are interconnected on a network to interact (\cite{nakamaru1997evolution,nowak1992evolutionary}) seems to provide a solution. Indeed, structuring the interactions increases both realism of models and the realism of conclusions allowing the survival of cooperation in the population. 
When considering structured interactions, the impact of network topology on the diffusion of cooperation needs can be addressed (\cite{Virtuallabs,santos2005scale,johnson2003social, ohtsuki2006simple}) and the realism of model can be improved allowing the interaction structure to co-evolve with agent strategies. In this case, chances for cooperation are enhanced (\cite{santos2006cooperation,yamagishi1996selective,yamagishi1994prisoner}). More specifically, among the mechanisms that improve the conditions of cooperation in dynamic networks are the possibility of parter selection, exclusion of defecting agents, and exit from relationships (\cite{schuessler1989exit,vanberg1992rationality,yamagishi1996selective}).

A recent series of our (\cite{righi2014emotional,righi2014parallel}) and other authors' papers (\cite{szolnoki2013evolution}), extended the analysis of the emergence of cooperation to signed networks. We introduced the possibility of network ties to turn positive or negative, or to be deleted and relinked as a consequence of previous interactions. We showed that the presence of emotional strategies - that use the {\it emotional} content implied by the relational signs in social interactions when considering the strategy to play - is pivotal for the survival and diffusion of cooperation. Indeed, in some cases, this strategy acts as a catalyst for unconditional cooperation rather than gaining dominance itself. We characterized the conditions in terms of the speed of evolution and selection pressure that allow the emergence of cooperation. In line with the literature, we found that relatively low rates of strategy adoptions and high rates of rewiring of stressed links are required in order to sustain cooperation. 

In this paper, we further extend the study of the emergence of cooperation in signed networks studying the impact of variance (or heterogeneity) in the number of connections of agents at the outset. In particular, we compare the results obtained on a regular lattice with those obtained on a comparable random network. We show that the increased degree variance at the outset extends the parameters' range under which cooperation survives in the long run. 
In this sense, our results confirm and extend those of \cite{santos2005scale}, and show that networks with high heterogeneity in degrees improve the conditions for the emergence of cooperation. 
Moreover, we show that the benefits in terms of increased space for cooperation by introducing the emotional strategy extend to both random networks and regular lattices.

In the remaining of this paper, we proceed as follows. First we discuss the characteristics of the agent-based model, then we report our results, and conclude with a brief discussion. 

\section{Model}

We consider a population of size $N$, connected by an undirected and non-weighted signed network. We restrict our interest to networks that are single components.
Each agent $i\in\{1,2, ..., n\}$ plays the single-shot Prisoner's Dilemma (PD) with each of his current neighbors, i.e. with a subset of the whole population $\mathcal{F}_i^t \subset N$. The cardinality $k_i^t$ of $\mathcal{F}_i^t$ is the degree (or number of network contacts) of the agent $i$, at time $t$. The network is signed and each tie is labelled either \textit{negative} or \textit{positive}.

We assume that the social network constrains the possible interactions so that only currently connected agents can play the game together. The payoff structure of the PD is reported in Table \ref{PDPayff}. When two agents cooperate with each other, each gets a reward (R). When they both defect, they are both punished (P). When one agent defects and the other cooperates, the first gets a temptation payoff (T), while his partner obtains the sucker payoff (S). The PD is defined with payoffs $T>R>P>S$. A typical additional assumption, that we adopt here, is $T+S>R+P$ (\cite{axelrod2006evolution}). 

\begin{table}[h!]
\centering
\caption{The Prisoner's Dilemma payoff matrix. The numerical payoffs used here are the same of \cite{axelrod2006evolution}.}\label{PDPayff}
$\begin{array}{|p{0.7cm}|c|c|}
\hline
& \textrm{C} & \textrm{D} \\
\hline \textrm{C} & (R=3,R=3) & (S=0,T=5) \\
\hline \textrm{D} & (T=5,S=0) & (P=1,P=1) \\
\hline
\end{array}$
\end{table}

Agents play cooperation or defection in the PD according to their type. We consider three possible strategy types:
\begin{itemize}
\item Unconditional Cooperation (UC) that always cooperates, without taking into account the sign of the tie he shares with his interacting partner.
\item Unconditional Defection (UD) that always defects. 
\item Conditional Action (COND) that cooperates with agents he shares a positive tie with and defects with those he has a negative tie with\footnote{The opposite strategy, that of defecting with cooperators and cooperating with defectors is not considered as deemed to be unrealistic.}. We label this strategy as \textit{emotional}, because it is a trigger response to the valence of the relation.
\end{itemize}

We let our agent based model to run in time steps. Steps are iterated until an equilibrium is reached. The conditions for considering one configuration as an equilibrium are stringent. It is required that: (1) a transitory period of 150 steps has passed from the beginning of the simulation (2) in five randomly chosen periods of time since (each time has a probability $0.1$ to be selected) the configuration of both relational sign, network topology and agent types needs to be precisely the same. \footnote{Robustness checks with alternative parametrizations have been performed and they do not influence the results.} Each time step (say $t$), a set of actions are performed by each agent, with the updates being done in parallel. Agents interact with peers they were connected with at the previous time step ($t-1$) and eventual updates in signs or network topology are observed by partners only in the following step $t+1$. Following a typical implementation of the literature, we assume that each agent plays the PD with all agents in his first order social neighborhood (i.e. with each $j\in\mathcal{F}_i^{t-1}$) and the average payoff is used when updating the agent strategy. 
The interested reader can find in \cite{righi2014parallel} a discussion of the effects of using an alternative, sequential, updating protocol.

The dynamics of our model allows for the co-evolution of network signs, agent strategies, and network structure. At each time step, network signs and agents behavior influence each other and the latter also affects the evolution of network topology. More in detail, after each dyadic interaction, stressed network signs are updated (with probabilities $P_{neg}$ and $P_{pos}$) or deleted and substituted with a new one with a certain probability ($P_{rew}$). At the end of each time step, when all payoffs are calculated, agents update their strategy to one that has been more successful in their neighborhood, with a certain probability $P_{adopt}$ (see Algorithm \ref{parallelalgo}).

\begin{algorithm}[!ht]
% \SetLine % For v3.9
 \SetAlgoLined % For previous releases [?]
 \For{each agent $i$}
 {
 	Compute its social neighborhood $\mathcal{F}_i^{t-1} \in N$\;	
	 \For{each agent $j \in \mathcal{F}^{t-1}_i$}
	 {
		Play the PD and compute payoffs\;
		Update the relational sign between $i$ and $j$\;
		If tense, delete the link between $i$ and $j$ (with Probability $P_{rew}$)\; 
	}	
	Compute average payoff of agent $i$\; 
}	
\For{each agent $i$}
{
 	Observe the average payoffs of each agent $j \in {F}_i^{t-1}$\; 
	Adopt the strategy of one agent with (strictly) higher payoff (with probability $P_{adopt}$)\; 
}
\caption{Intra-step dynamics, repeated at each time step $t$. Details are provided in the next paragraph.}
\label{parallelalgo}
\end{algorithm}

Let's discuss each of the elements described in Algorithm \ref{parallelalgo} more in detail. 

{\bf Update of the relational sign between $i$ and $j$.}
 After each dyadic PD game, agents might update their relational sign with each other. Given the nature of the PD, there are three possible situations:
\begin{itemize}
\item {\it Both players cooperate}. In this case an existing positive connection remains positive, while a negative one turns positive.
\item {\it Both players defect}. Similarly, an existing positive relation is turned negative and a negative one remains so.
\item {\it One agent cooperates and the other defects}. In this case, the emotional content of the relationship is subject to stress. We assume that if the link is positive, then the cooperator is {\it frustrated} to have a positive relation to a defector. Therefore, we assume that the valence of the tie can turn negative with probability $P_{neg}$. If the link is negative, the defector might be interested in turning it into a positive tie. We assume that it happens with probability $P_{pos}$. There are two possible justifications for such behaviour. The first is that the defector feels remorse or moral guilt (as suggested by \cite{gaudou2014moral}). The second is instead purely selfish. The defecting partner is content to remain friends with the cooperator. This type of relationship provides him with a strictly higher payoff, in case he is paired with a COND player, whose action is sensitive to the sign of their relationship. It is logical to assume, however, that the frustration from the cooperator is larger, therefore we impose $P_{neg}>>P_{pos}$. \footnote{For the runs reported here, we fixed $P_{neg}=0.2$ and $P_{pos}=0.1$. These values are assumed equal for all agents. We run a sensitivity analysis of this parameter in \cite{righi2014emotional}}
\end{itemize}

{\bf Delete the link between $i$ and $j$ and create a new tie.} 
An agent, frustrated by the current behavior of the partner, may decide to delete the social connection completely with probability $P_{rew}$. In this sense, our network topology co-evolves with agents' strategies endogenously (similarly to \cite{santos2006cooperation}). $P_{rew}$, called rewiring probability, is assumed to be equal for the whole population and it non-strategic. When rewiring takes place, once the old link is erased, a new one is created with another agent. In line with the sociological literature (\cite{granovetter1973strength}), we assume that there is a tendency towards transitive closure.\footnote{The assumption of existence of transitive closure makes the model more realistic and increase cooperation. As shown in \cite{righi2014emotional} however, our results are qualitatively robust when we relax this assumption and consider totally random rewiring.} New connections are created to friends of friends (excluding the possibility of connections to friends of enemies, to enemies of friends, or to enemies of enemies). In order to introduce some social noise, with a probability $P_{rand}$, rewiring takes place to a randomly selected agent in the population. \footnote{This parameter is assumed to be small but positive. Its value is fixed to $P_{rand}=0.01$ in our simulations.} The network structure evolves dynamically through rewiring. This implies that, while the initial topology is either a regular lattice or a random network, it does not necessarily remain of this type  - and in general, it does not.

{\bf Adopt a better strategy.}
 Agents observe their average payoffs as well as the ones of the agents in their social neighborhood, and are thus able to measure the relative local efficiency of their strategy. If a subset of agents in $\mathcal{F}_i^{t-1}$ has a payoff  at time $t$ higher than his own, then agent $i$ will adopt the strategy played in $t$ by one of them, selected uniformly at random. Evolutionary update happens, for each agent, with probability $P_{adopt}$ which is assumed to be equal for all agents.

\begin{figure}[t]
\centering
\includegraphics[width=0.48\textwidth]{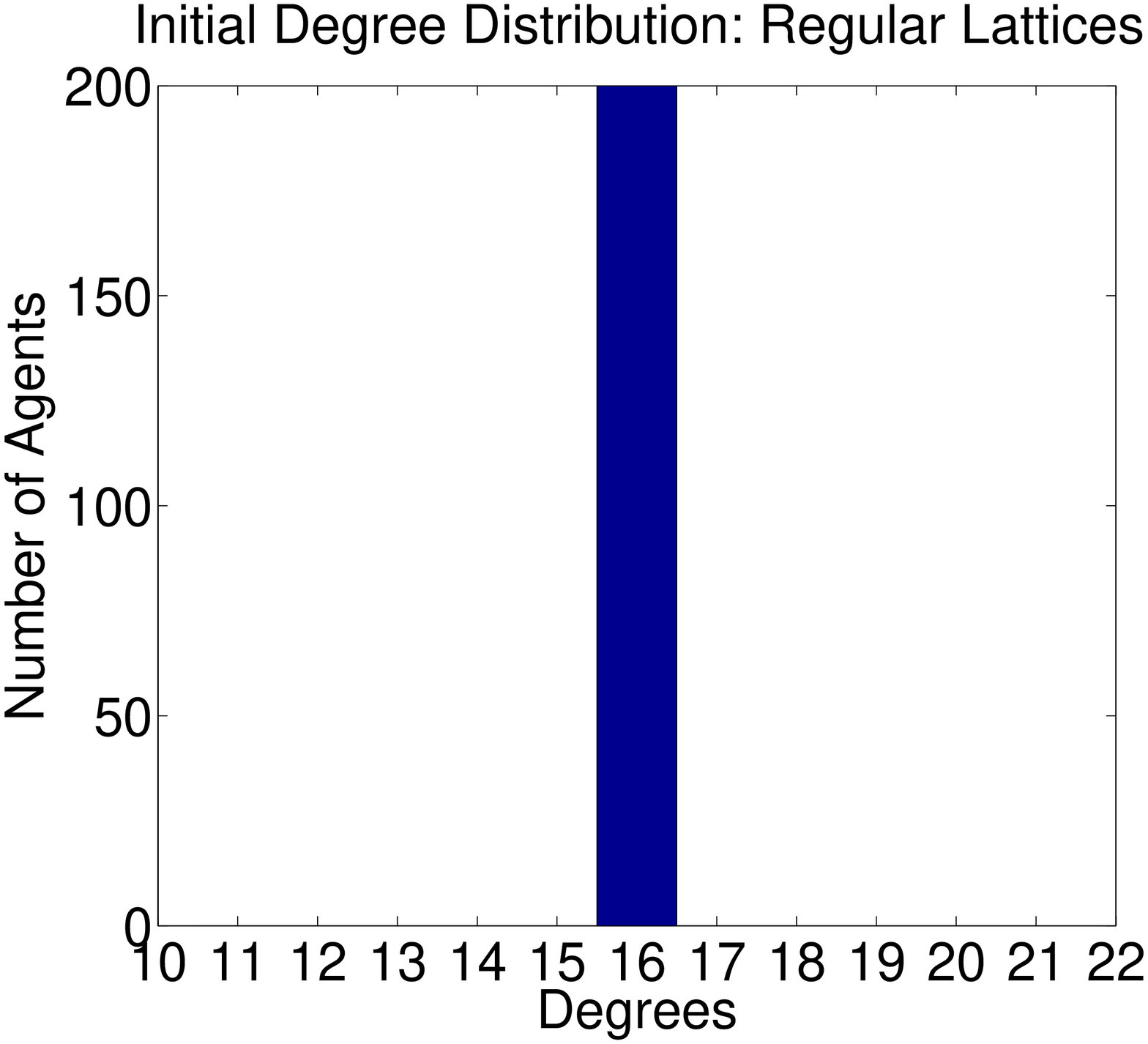}
\includegraphics[width=0.48\textwidth]{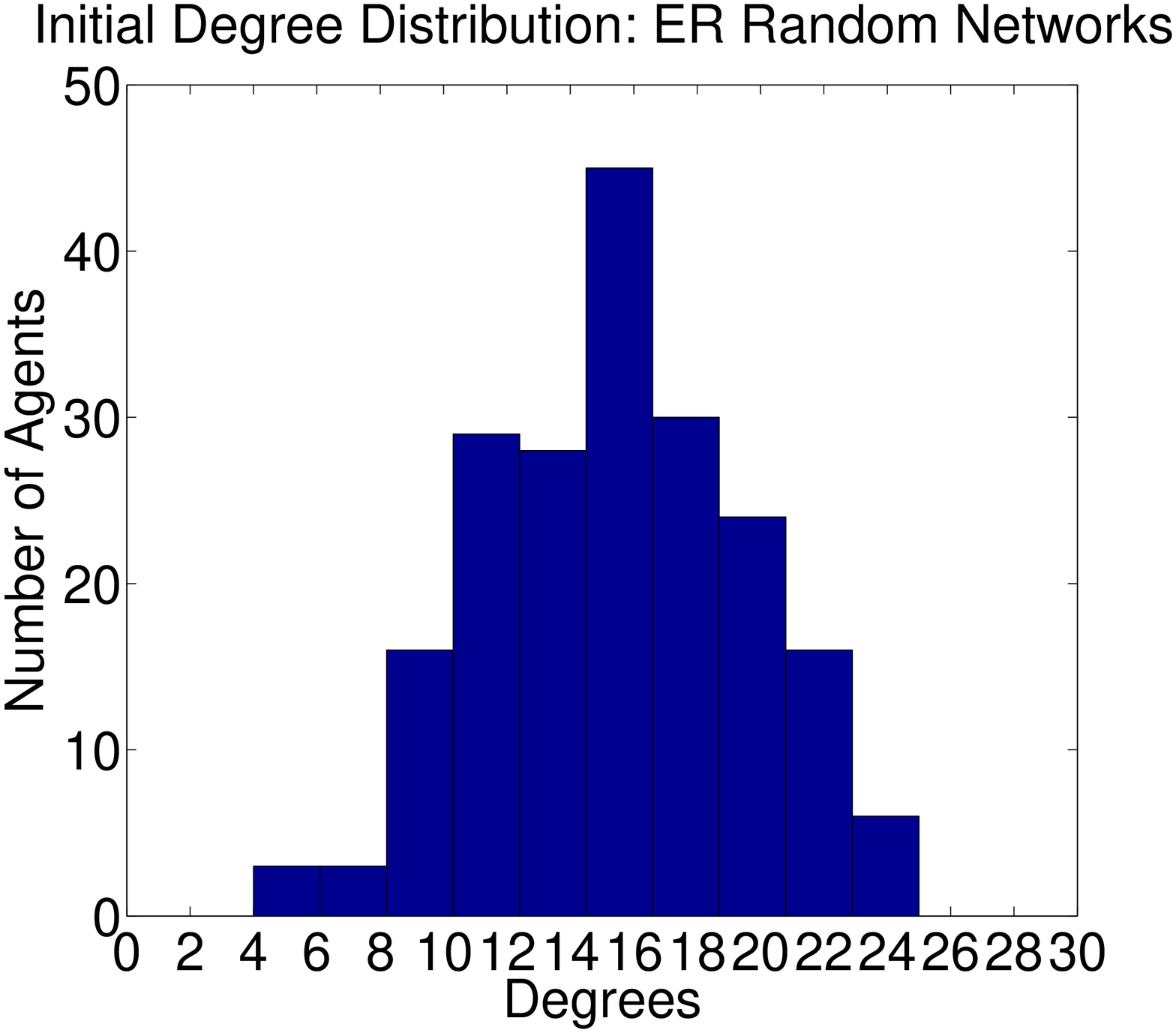}
\caption{Degree distributions in typical networks used for initialize our simulations. In both the regular lattice case (Left Panel) and the random network case (Right Panel) the average number of connections per agent is 16. $N=200$.}
\label{distros}
\end{figure}

{\bf Simulations Calibration.} 
Concerning the initial structure of the social network, we provide results for two cases. In a first set of simulations we assume that agents are laid on a {\it regular lattice} in which every agent has precisely 16 connections (the degree distribution is therefore degenerate as shown in the Left Panel of Figure \ref{distros}). Then we introduce heterogeneity in degree distribution and we study networks initialized as Erd\H{o}s-R\'{e}nyi (\cite{erdHos1959random}) {\it random graph} (an example of the resulting degree distribution is provided in the Right Panel of Figure \ref{distros}). In order to make the results comparable we impose that each pair of nodes is connected with an independent probability $P_{link}=0.16$ so that the degree distribution is centered around 16 with a standard deviation of about 4.
Moreover, agents are assigned with one of the three strategies randomly in equal proportions. In the absence of conditional players, the proportion of UDs and of UCs are $1/2$. When CONDs are added, then the starting proportion of each type of agent is $1/3$. Finally, the relational signs are randomly distributed and initialized so that each link has a 50% chance of being either negative or positive. \footnote{Results for different initial populations are discussed in the companion paper \cite{righi2014emotional}.}

\section{Results}

Our aim with this study is to characterize the parameter configurations that favor the evolution of cooperation in dynamic signed networks. We focus on the effect of conditional (or emotional) strategy in two different network initializations: in a regular lattice and in an Erd\H{o}s-R\'{e}nyi  random graph. The two main dynamic forces that operate in our model are the evolutionary pressure ($P_{adopt}$) and the network update dynamics ($P_{rew}$). Our strategy is to analyze their impact, changing their relative strength progressively. For each possible combination of the two probabilities (each studied for values between $0$ and $1$ with a granularity of $0.05$) we show results concerning the average proportion (calculated in 50 simulations) of the agents and network ties surviving at the steady state.~\footnote{Standard deviations are not reported here and are available upon request. The variability of the results is quite small except in the area of the phase transition between the configurations in which cooperation survives and those where it disappears completely. Only statistically significant phenomena are studied and discussed in the following.} 

\begin{figure*}[t]
\centering
\includegraphics[width=0.32\textwidth]{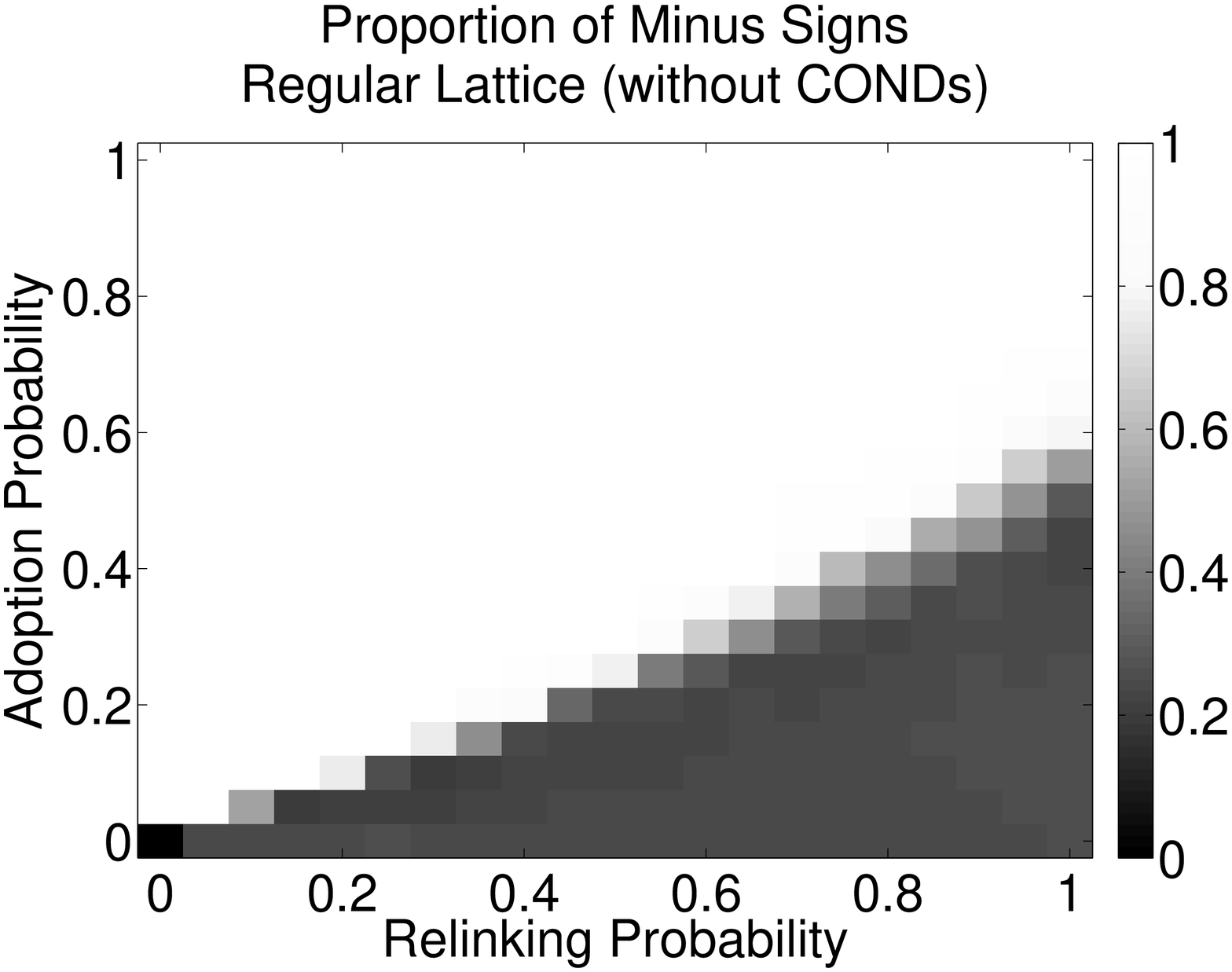}
\includegraphics[width=0.32\textwidth]{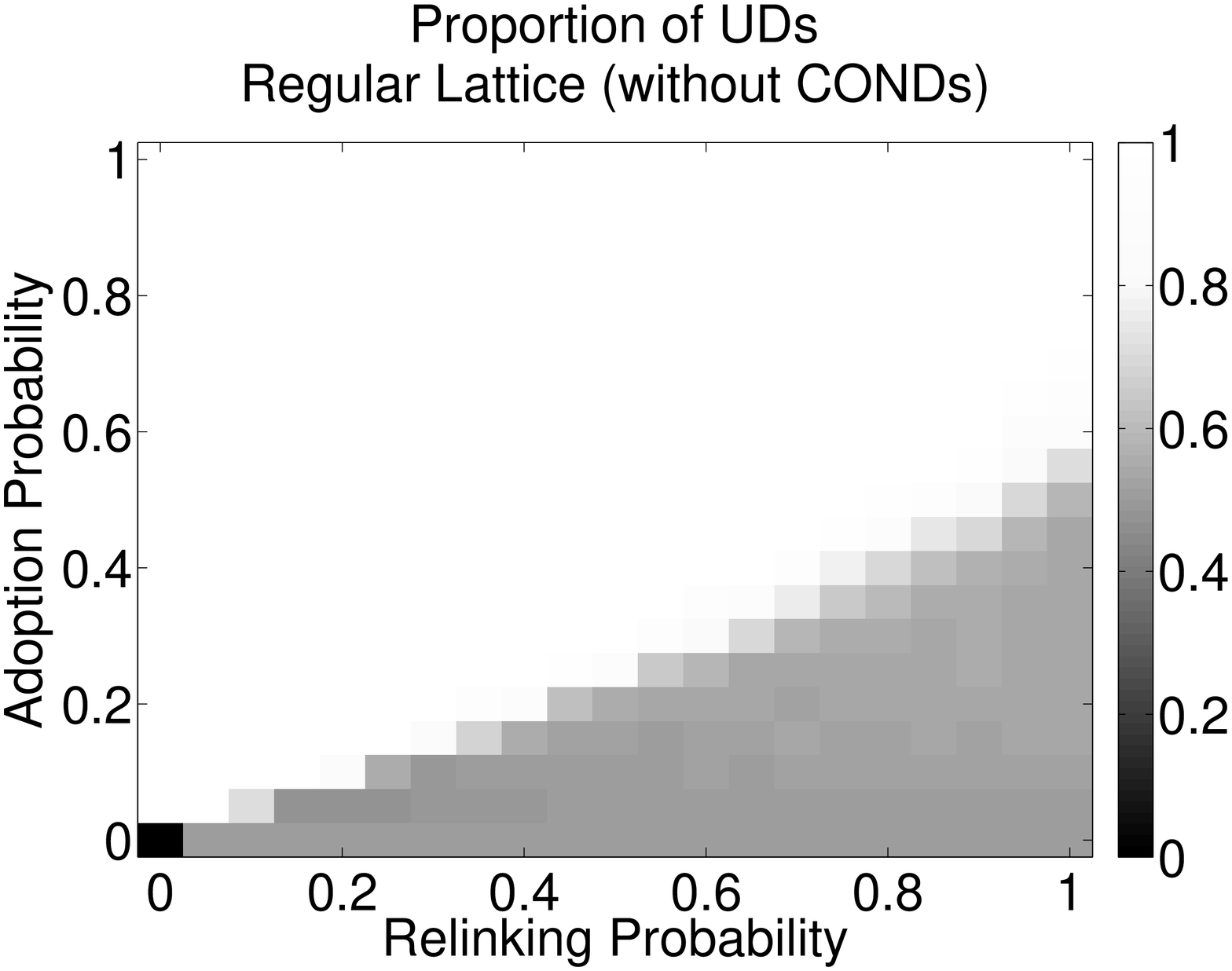}
\includegraphics[width=0.32\textwidth]{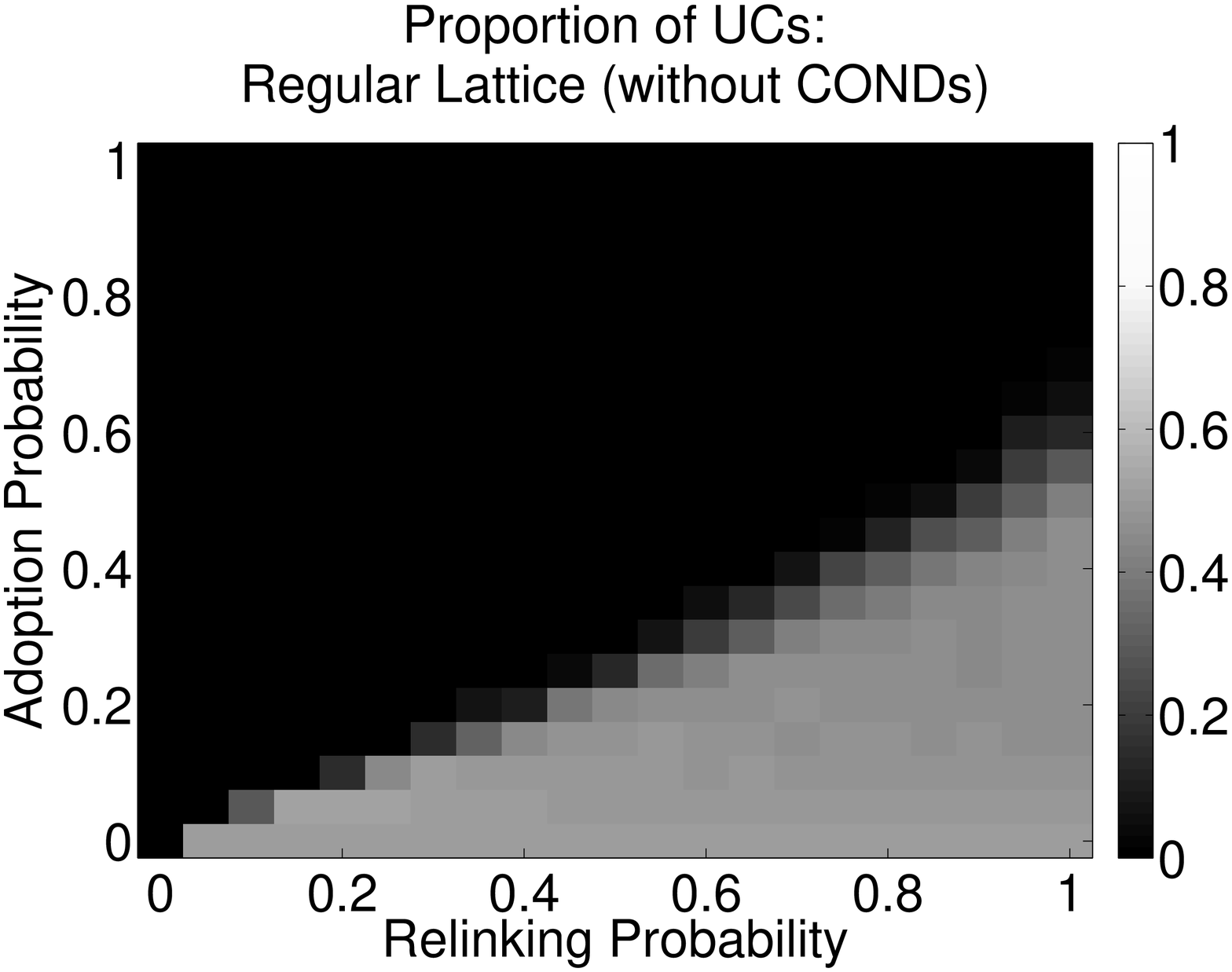}\\
\includegraphics[width=0.32\textwidth]{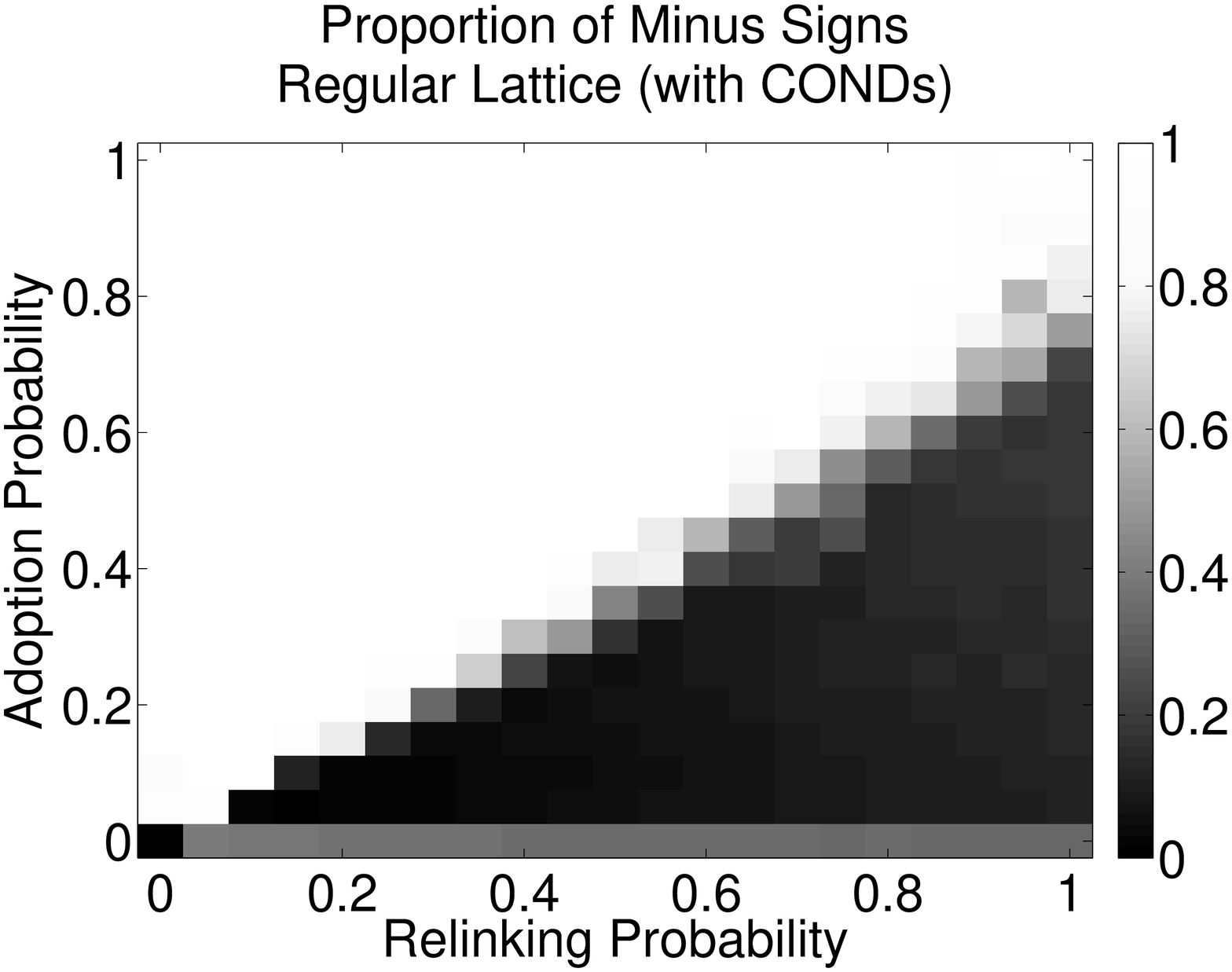}
\includegraphics[width=0.32\textwidth]{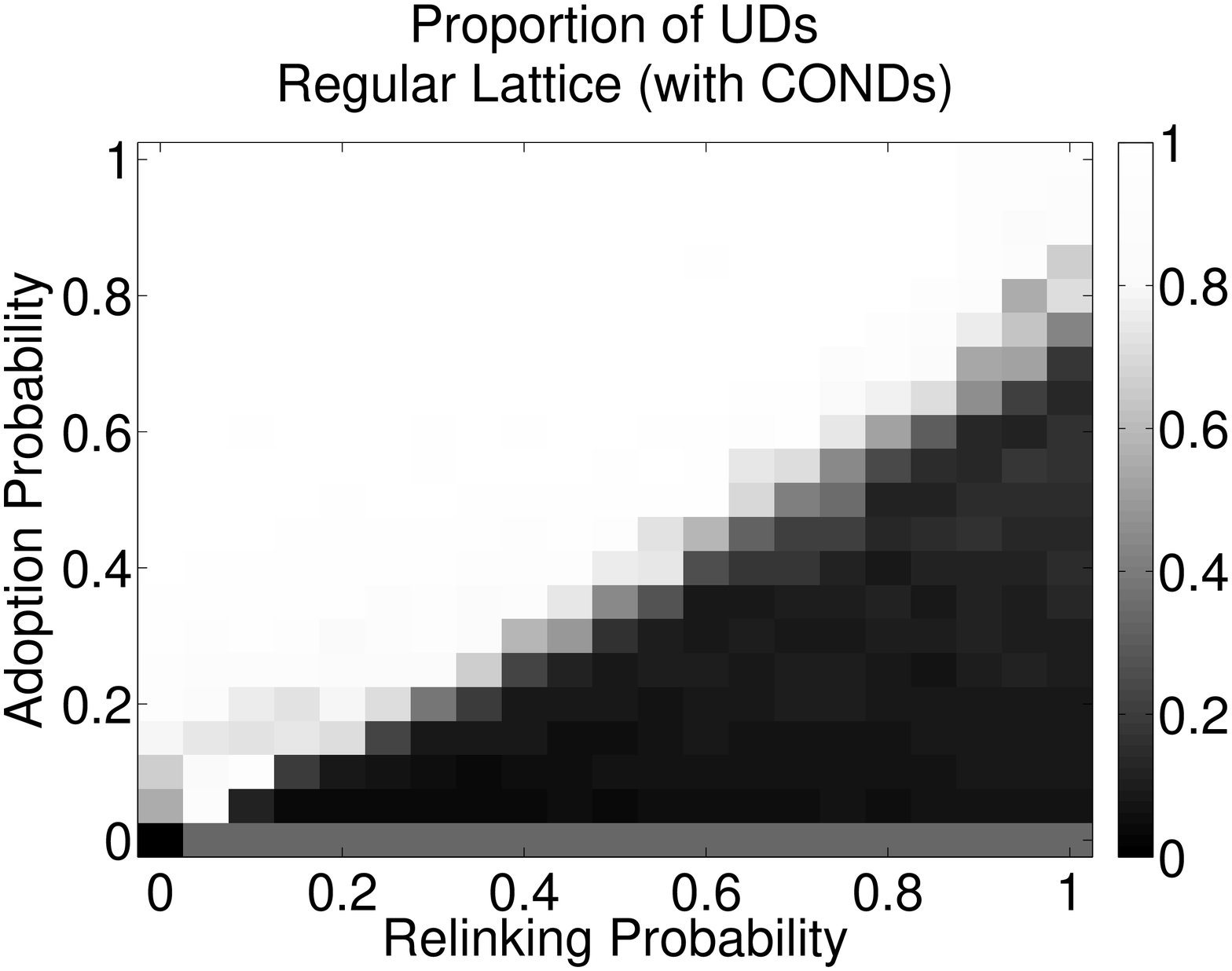}
\includegraphics[width=0.32\textwidth]{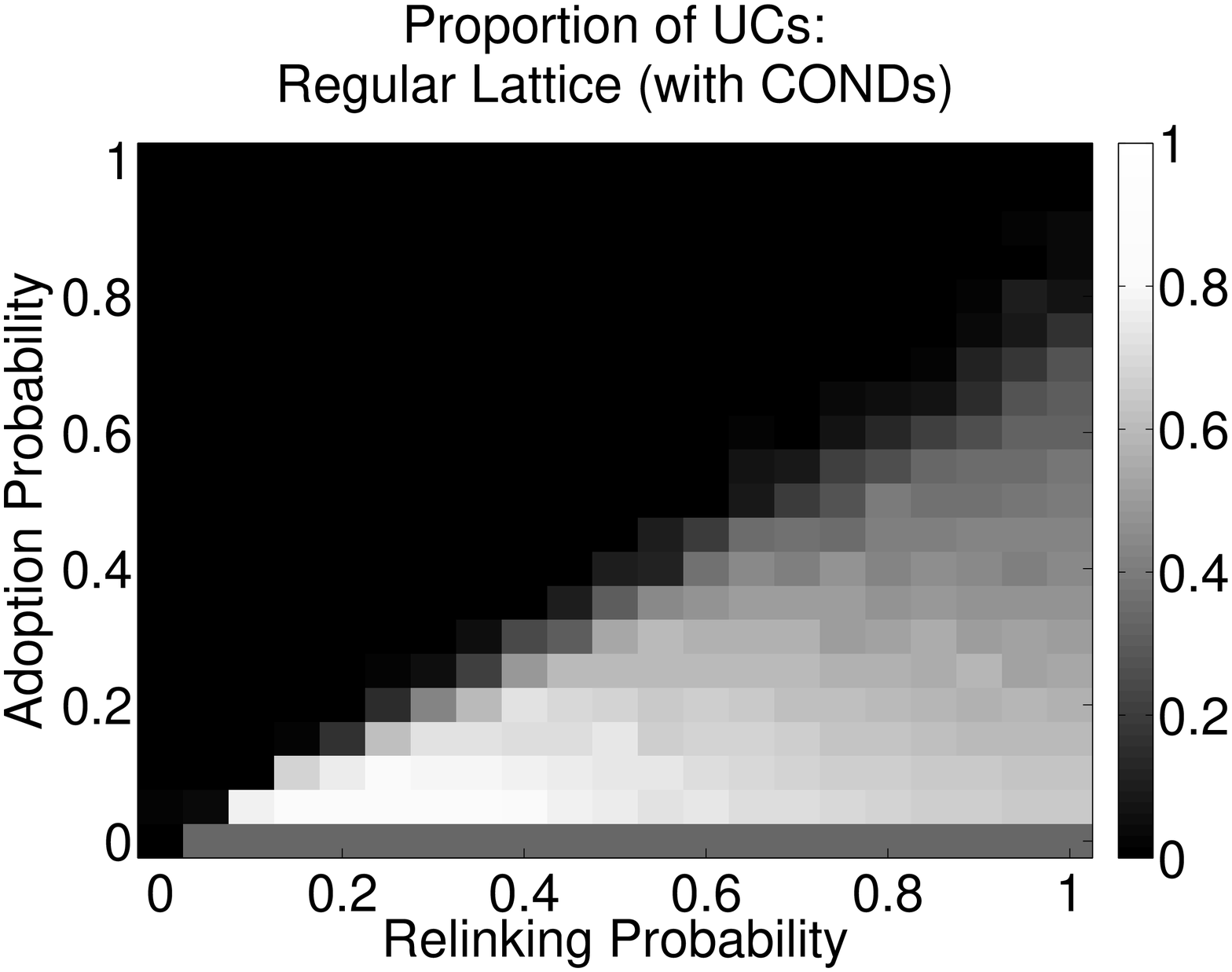}\\

\caption{Effect of the competing dynamics of strategy adoption (vertical axis) and of rewiring of stressed links (horizontal axis) on the final proportion of negative ties in the network (Left Panels), of UDs (Central Panels) and of UCs (Right Panels). Top Panels show results for populations initialized as equally divided between UCs and UDs (where there are no CONDs). Lower Panels show results for populations initialized as equally divided among the three different agent types. The social networks are initialized as {\it regular lattices} where each agent has 16 connections. Each datapoint is the average of 50 simulations. For each simulations $N=200$ and network signs are randomly initialized with equal probability.}
\label{FixedPFlipRL}
\end{figure*}

\begin{figure*}[t]
\centering
\includegraphics[width=0.32\textwidth]{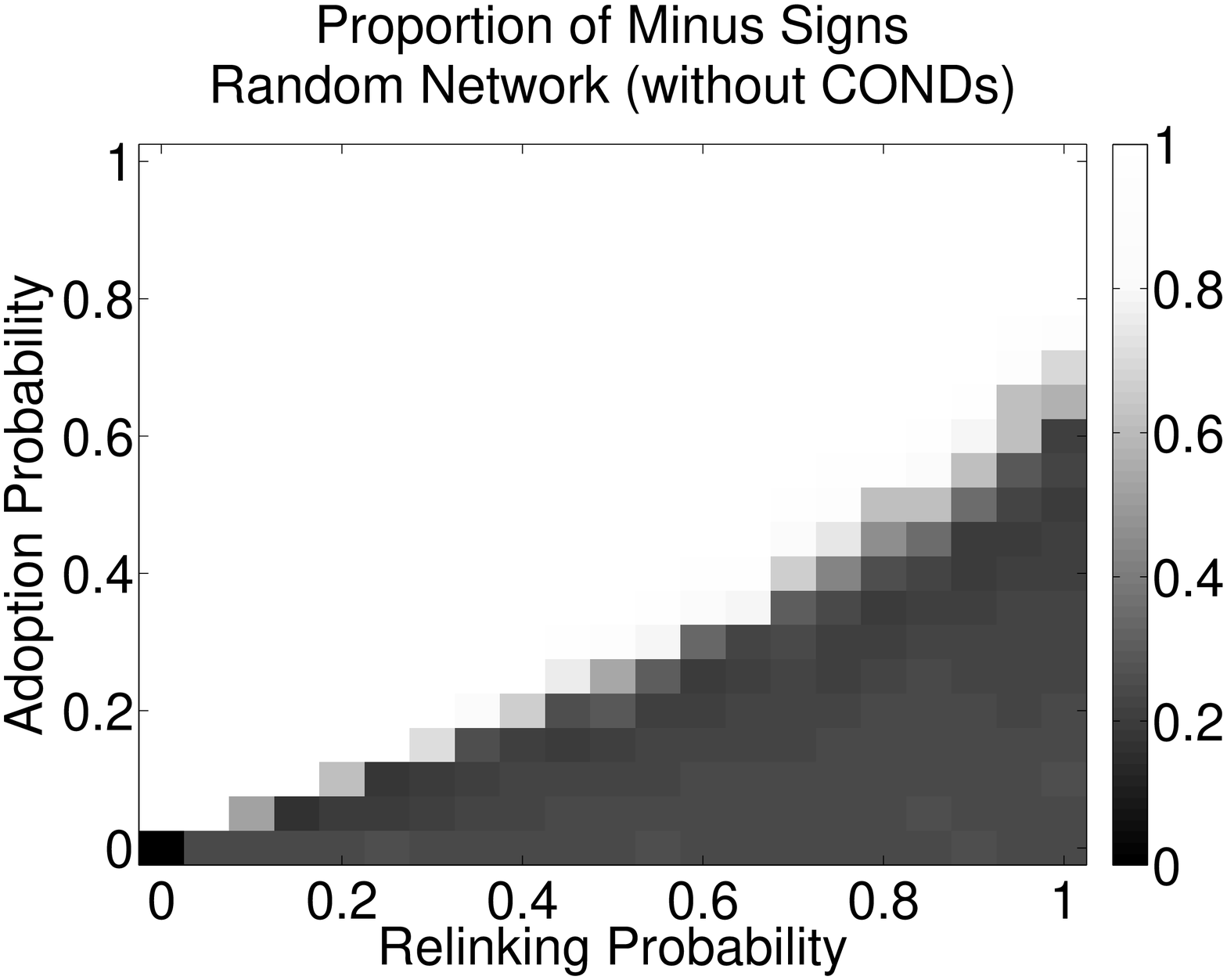}
\includegraphics[width=0.32\textwidth]{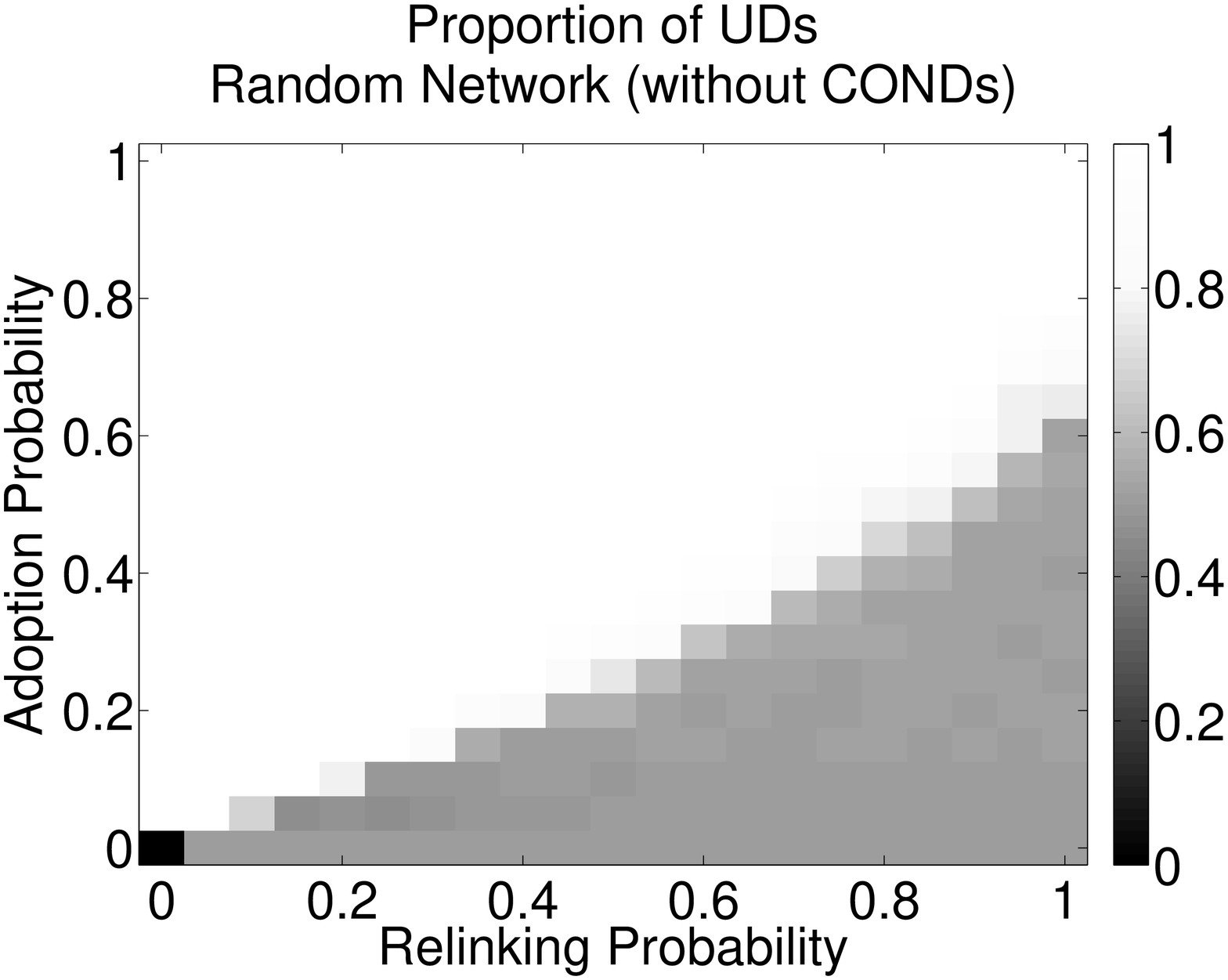}
\includegraphics[width=0.32\textwidth]{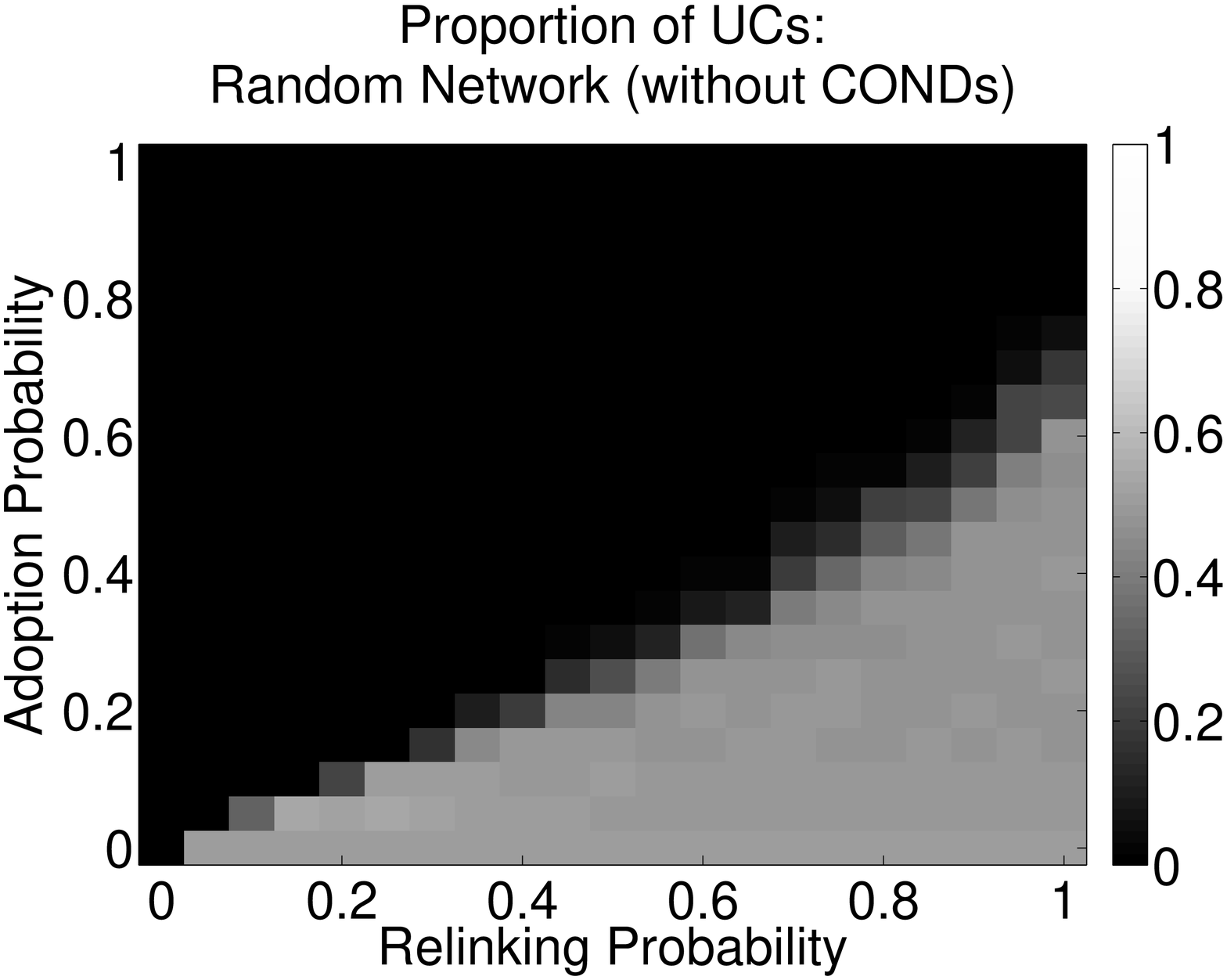}\\
\includegraphics[width=0.32\textwidth]{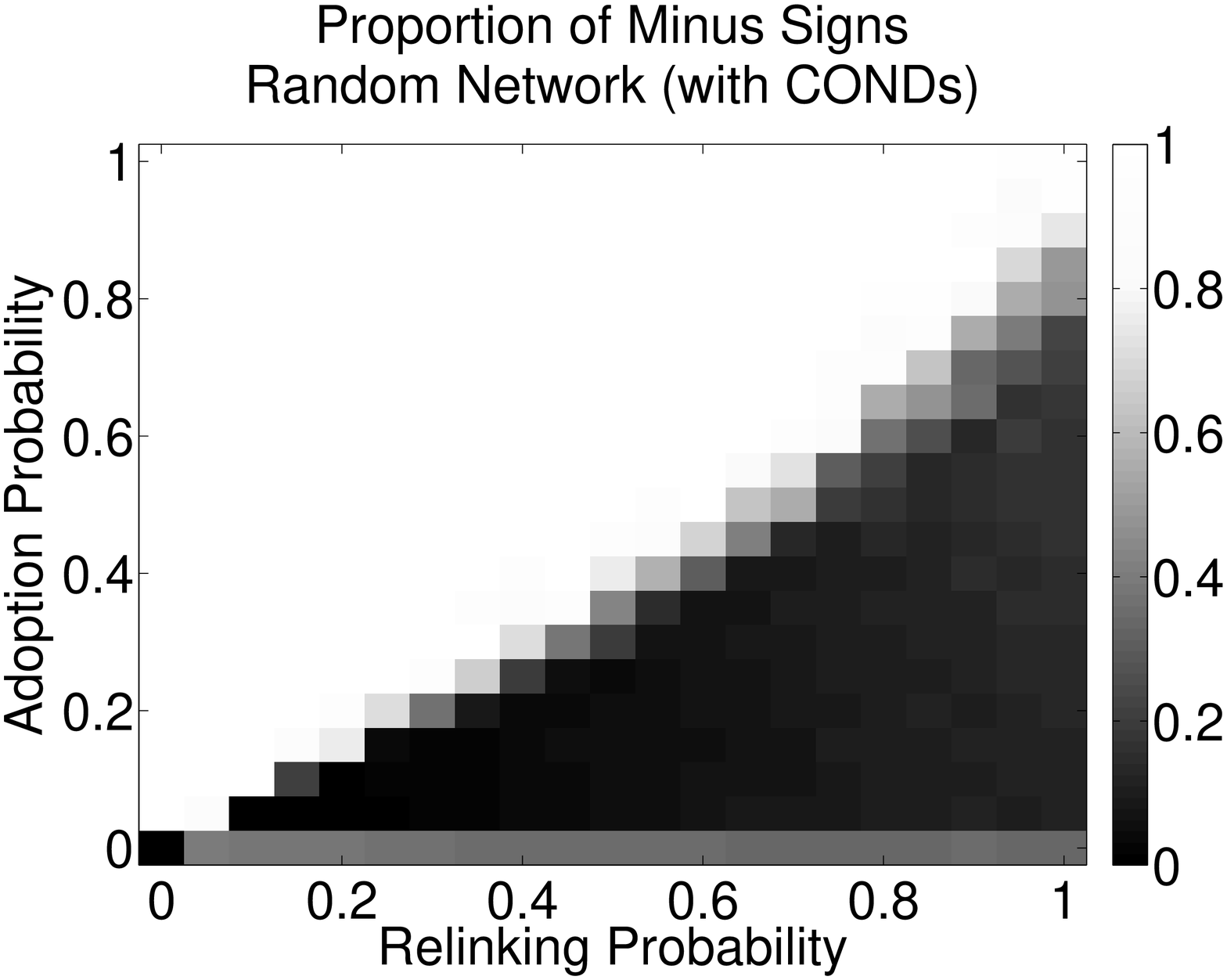}
\includegraphics[width=0.32\textwidth]{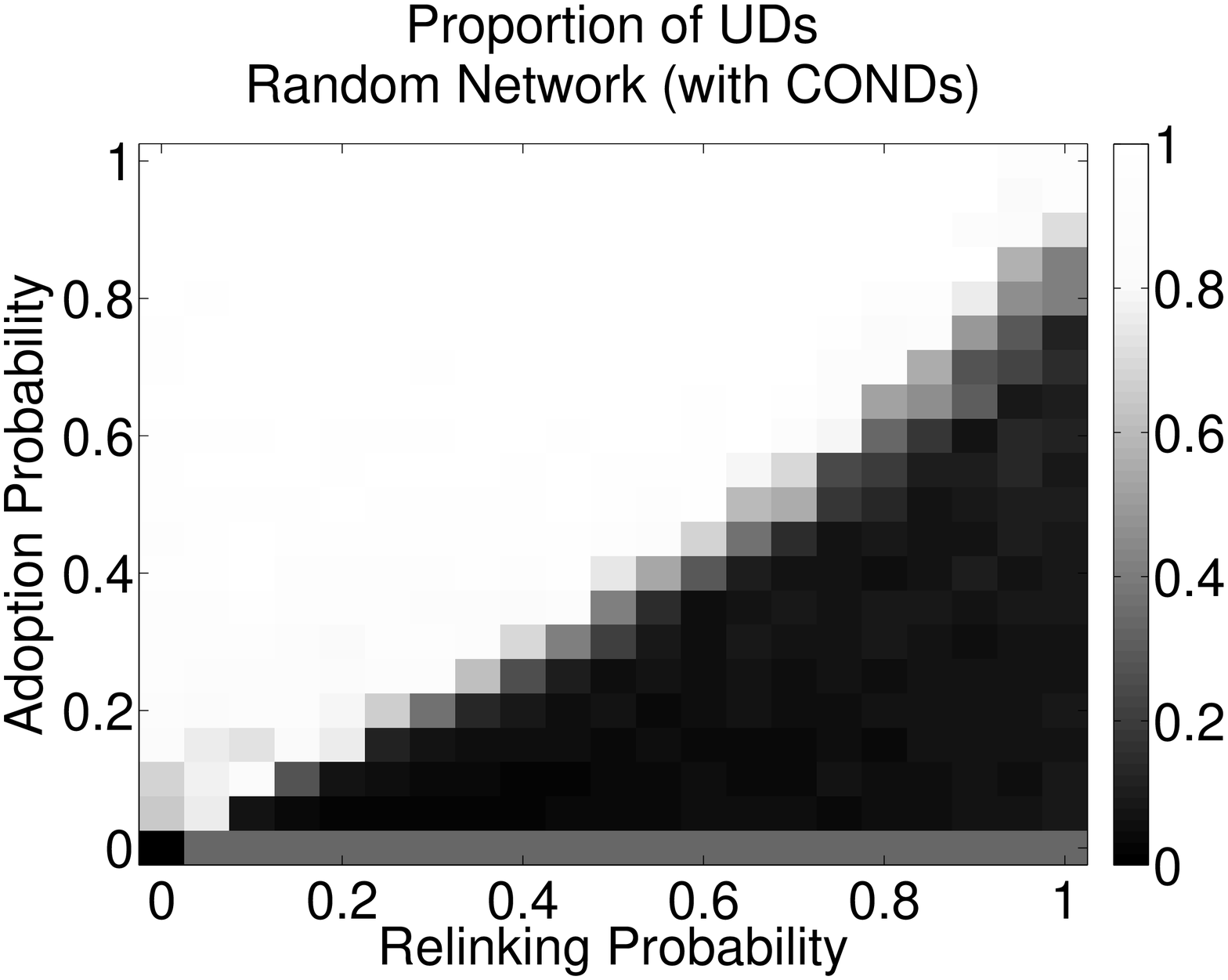}
\includegraphics[width=0.32\textwidth]{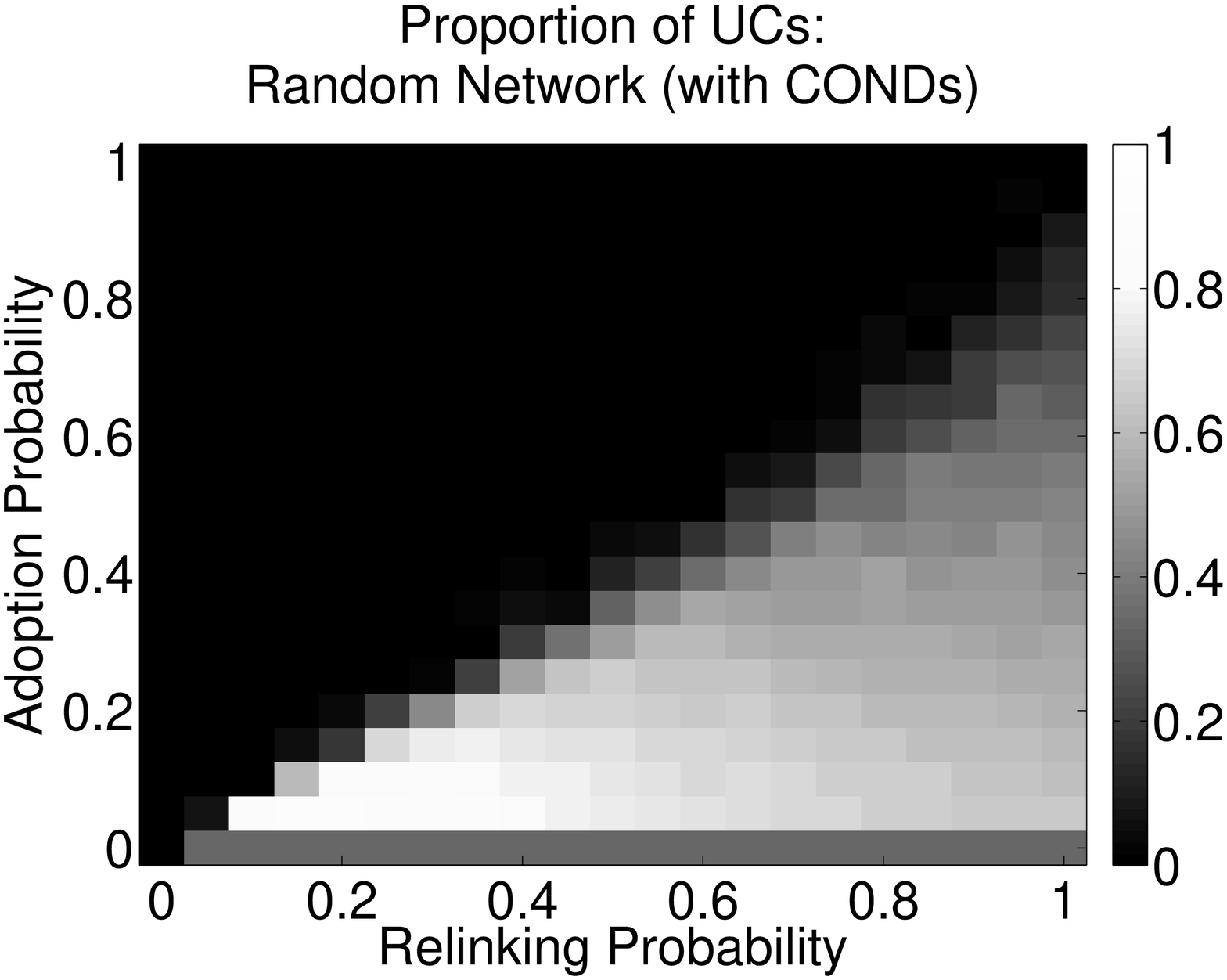}\\
\caption{Effect of the competing dynamics of strategy adoption (vertical axis) and of rewiring of stressed links (horizontal axis) on the final proportion of negative ties in the network (Left Panels), of UDs (Central Panels) and of UCs (Right Panels). Top Panels show results for populations initialized as equally divided between UCs and UDs (where there are no CONDs). Lower Panels show results for populations initialized as equally divided among the three different types of agents. The social network are initialized as Erd\H{o}s-R\'{e}nyi {\it random network} with $P_{Link}=0.16$. Each datapoint is the average of 50 simulations. Network signs are randomly initialized with equal probability. $N=200$. }
\label{FixedPFlipRand}
\end{figure*}

Figures \ref{FixedPFlipRL} and \ref{FixedPFlipRand} report results for two alternative cases each. 
In Figure \ref{FixedPFlipRL}, the social network is initialized so that every agent has initially the same amount of connections, which defines a regular lattice. In Figure \ref{FixedPFlipRand}), results are shown for a setup where degree variance is introduced and the network is initialized as an Erd\H{o}s-R\'{e}nyi random network. 
In the Top Panels of the Figures, the population is initialized as equally divided between unconditional cooperators and unconditional defectors. This simulation is compared with the case (Lower Panels), in which the population is initialized as divided equally among UCs, UDs and CONDs. 

One can observe several similarities in the results from the two kinds of starting configurations. As noted in our previous contributions (\cite{righi2014emotional,righi2014parallel}), and coherently to what observed by \cite{santos2006cooperation}, the relative speed (i.e. the probability) of the network topology update and of strategy adoption have two opposite effects on the viability of cooperative strategies. Increasing the speed of adoption of better strategies favors defection, as this is the strategy that maximizes payoffs in dyadic terms. At the opposite, a relatively high degree of network updating leads to higher proportions of cooperation, as it helps the formation of clusters of cooperators. 

From Figures \ref{FixedPFlipRL} and \ref{FixedPFlipRand}, we can observe that defectors suddenly lose dominance when a certain ratio between the two dynamic forces is reached. In the case of the regular lattice initialization without emotional strategies, the cooperation survives if the approximate relation $P_{rew} > 2 P_{adopt}$ holds. The chances of cooperation are increased for ER networs compared to a regular lattice initialization for any combination of $P_{rew}$ and $P_{adopt}$. In this case, the condition for cooperation to survive is $P_{rew} > 5/3 P_{adopt} (\sim 1.6 P_{adopt})$. 

Let's speculate about the reason for this improvement. In the absence of CONDs, the only force preventing UCs from being eliminated from the network is the rewiring of stressed ties. As we discussed, this process tends to create clusters of cooperators that can then survive. When all agents have the same connectivity, they all require similar amounts of time to rewire their connections to UDs, which can then spread locally and dominate the final population. When the network is initialized as random, some of the agents have less connections than the average, and they become isolated from the defector faster; substituting negative stressed connections with positive ones via transitive closure. The new connections are more likely to be with CONDs (when present) or UCs (which cooperate at least when given the opportunity and thus tend to develop positive ties) than with defectors, given the positive relations involved. These agents constitute therefore the nucleus of cooperative clusters around which more connected cooperative peers can survive. 

In summary, degree heterogeneity provides {\it time} for clusters to form, even when the ratio between $P_{rew}$ and $P_{adopt}$ is less favorable.
The positive effect of the increased heterogeneity for cooperation is stronger in more dynamic networks (higher $P_{rew}$) since agents with few connections extricate more efficiently their leverage effect on the formation of cooperative clusters.
Introducing a variability in the degrees of the agents, thus increases the range of parameters in which cooperation survives and diffuses in the population. This result confirms the one obtained by \cite{santos2005scale}. We consider, however, a dynamic environment in which agent strategies co-evolve with relational signs and network topology. Moreover, from the purely topological point of view, we show that cooperation can be increased through heterogeneity also without recurring alterations of the randomness of the network (such as preferential attachment or network growth). 

In both types of network initializations, when the conditions for cooperation to survive are met and CONDs are absent, the results for different parameter combinations are rather similar. They indicate that about 25\% of the signs in the network are negative and the population turns out to be equally split between UCs and UDs. Regardless of the starting network, we observe that when cooperation survives, it does not diffuses. In both cases, the proportion of cooperators remains similar (or just below) its initial setup value. We can understand this result observing that the sole driving force allowing the survival of cooperation (in the absence of emotional strategies) is the rewiring of stressed links. In this sense, $P_{rew}$ has a purely positive effect on cooperation and $P_{adopt}$ has a purely negative one. When the first dominates, cooperation survives, when the second dominates cooperation disappears; hence there is the sharp phase transition between the two states. 

As noted in our previous work (\cite{righi2014emotional}), the introduction of the COND strategy relaxes the parametric conditions for cooperation to survive. In the context of this paper, we note that this happens both for the regular lattice initialization, where the approximate condition for the survival of cooperation becomes $P_{rew} > 20/13 P_{adopt} (\sim1.53 P_{adopt})$, and for the random network where it becomes $P_{rew} > 4/3 P_{adopt} (\sim 1.3 P_{adopt})$. The relative effect of introducing CONDs in the population is thus similar in terms of the proportion of parameters in which cooperation becomes viable and thus the two initializations can be discussed together.

\begin{figure}[h]
\centering
\includegraphics[width=0.48\textwidth]{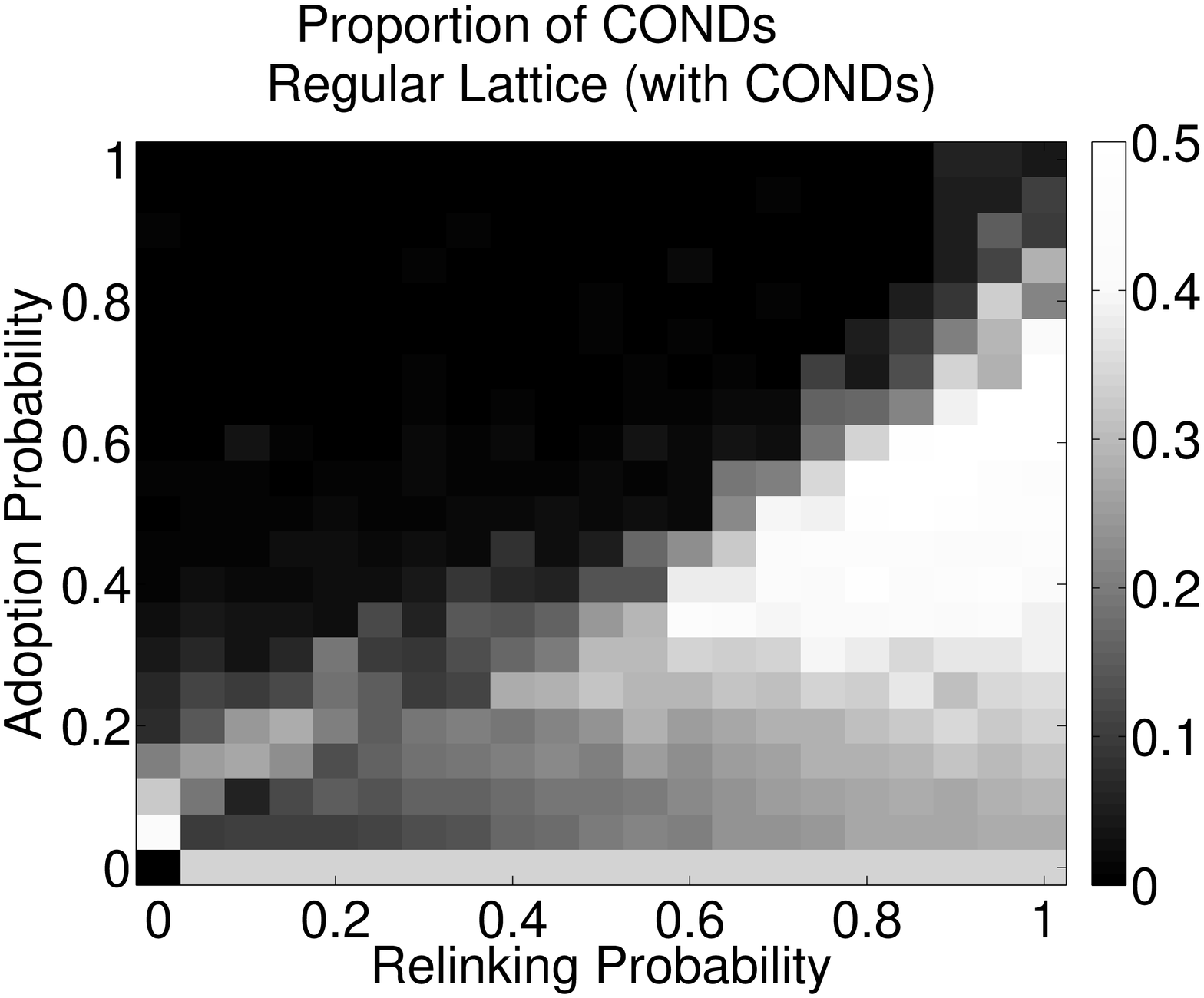}
\includegraphics[width=0.48\textwidth]{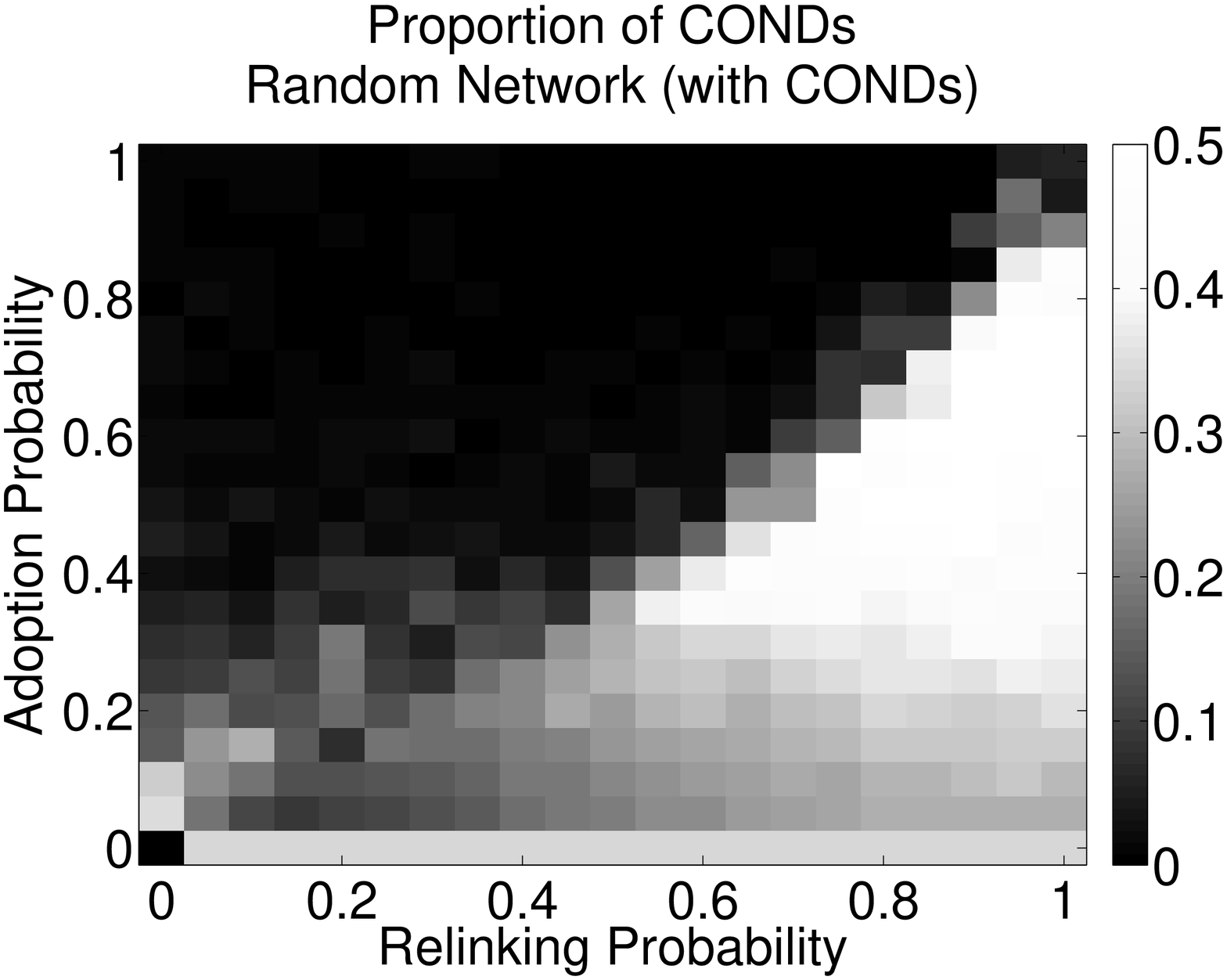}
\caption{Effect of the competing dynamics of strategy adoption (vertical axis) and of rewiring of stressed links (horizontal axis) on the final proportion conditional (COND) agents. Left Panel: results for the regular lattice initialization. Right Panel: results for the random network initialization}
\label{conds}
\end{figure}

Understanding this result requires a closer look at the final proportions of agents and relational signs in the area that allows the survival of cooperation. Here, the final proportion of UC agents ranges from 25\% to 75\% of the population, with this probability decreasing monotonically as the adoption and rewiring probability increases. Confronting the results regarding UCs with those regarding UDs, one can notice that the decrease in the proportion of UCs benefits UDs little (their proportion passes from a minimum of about 5\% to a maximum of 17\%, but much more the conditional players (see Figure \ref{conds}). 
In the same area, also the proportion of negative ties progressively increases from about 4\% to about 25\%, but never exceeds this value. We can thus conclude that, when cooperation is viable, clusters are formed in which CONDs and UCs are intertwined by positive links and therefore are functionally indistinguishable. Moreover, while in presence of conditional agents cooperative behaviour spread in the population, the dominant type of cooperation (conditional or unconditional) depends on the relative strength of our two main dynamic variables. 

For environments with relatively {\it low frequency} of network and evolutionary updates, the COND strategy acts as a shield for UCs. When in contact with both pure cooperators and pure defectors, agents playing this strategy tend to enjoy higher payoffs than those who always defect (COND gets payoffs from cooperation when interacting with the UC type, while avoiding the {\it sucker} position when interacting with UDs), and therefore tends to replace them due to the evolutive pressure. As the proportion of UDs decreases and segregation increases, pure cooperation becomes the optimal strategy, as it avoids "errors" due to the mis-interpretation of network signs. Thus, UCs tend to diffuse at the expense of CONDs, and the final proportion of unconditional cooperators tends to be high. 
At the opposite, when the two dynamic updates happen relatively fast, this dynamics reverse in favor of COND. When adoption of strategies with higher payoffs is faster, the cooperation is in general more difficult to sustain and pure defection diffuses more. Under these conditions, emotional agents, being able to discern among cooperators (with whom they tend to form positive ties) and defectors (with whom they tend to form negative ties) suffer less {\it sucker} payoffs from pure defection than pure cooperators; which therefore tend to disappear faster. In this more dynamic setup, the number of cooperators reduces too fast to regain dominance later, and conditional cooperation turns out as dominant.

\section{Conclusions}
The problem of evolution of cooperation has been widely studied in social sciences. Unlike most of the previous literature that considered only positive relations, we introduced negative ties as a force that is able to influence agents behavior. We presented results from an agent based model, where we studied the evolution of cooperation in dynamic signed networks in which agent strategies co-evolved with relational signs and network topology. Agents played the Prisoner's Dilemma with their current neighbors and the result of dyadic interactions drove the evolution of relational signs and network relations. The average performance of a strategy across all interactions of one individual was defined as the fitness value that determined the evolutionary process. 

In this paper, we performed an extensive simulation analysis of our model focusing on the effects on the survival of cooperative strategies as (1) network topology was varied, considering a regular lattice and a random network initialization; and as (2) a sign dependent strategy that considers the network signs when deciding whether to cooperate or to defect was introduced. 
We provided results for all possible combinations of the two main dynamic forces of this model by progressively changing both the probability of adopting more fitting strategies and the probability of rewiring tense connections.

In all cases, and in line with the literature, we showed that higher strategy evolution rates reduced the combinations of parameters in which cooperation survived (favoring the dyadic dominant strategy: to defect), while increasing the rewiring probability helped isolating cooperators from defectors thus favoring the survival of cooperation.

Random networks provided more place for the emergence of cooperative behavior for a larger set of parameters than regular lattices. This result is similar to the one of \cite{santos2005scale}, however our outcome follows from a different mechanism. In \cite{santos2005scale}, cooperation diffusion followed from the the presence of very connected hubs. In our setup, there were no such hubs (degrees have a bell shaped distribution around a characteristic degree and connections are purely random). The process allowing the diffusion of cooperation is the presence of individuals which are less connected than the average. By segregating early on in the simulations from defectors, these created the nuclei around which cooperative clusters could emerge. 

Extending the analysis of \cite{righi2014emotional} to regular lattices, we studied the effects of the introduction of a conditional strategy that considers the relational signs to the partner to decide whether to cooperate or defect. The conditional strategy  enlarged the parametric space in which cooperation evolved. Despite the advantage of being able to use more information, however, and regardless of the network topology adopted, the conditional strategy gained dominance itself only in a few, rapidly changing environment (where both adoption and rewiring happened relatively frequently). In these situations, the better performance of the COND strategy against pure defectors made the spread in the population possible. When the network and strategies were more stable, the conditional strategy acted instead as catalyst for the diffusion of unconditionally cooperative behavior.

The work presented in this paper is a first step in understanding the role of network topology in the diffusion of cooperation in dynamic signed networks. While a preliminary analysis has been introduced in \cite{righi2013signed}, further studies are required to address the issue of the evolution of cooperation in non-random signed networks systematically. In particular, more realistic network initializations, such as  scale-free and small-world networks, could be analyzed.

\section*{Acknowledgments}
The authors wish to thank the "Lend\"{u}let" program of the Hungarian Academy of Sciences and the Centre for Social Sciences for their financial and organizational support and three anonymous referees for their useful comments.  

\section*{Authors Biographies}

{\bf SIMONE RIGHI} is currently a Research Fellow at "Lend\"ulet" Research Center for Educational and Network Studies (RECENS) of the Hungarian Academy of Sciences. He studied mathematics and economics at the University of Namur (Belgium) where he obtained a Ph.D. in Economics in 2012 with a thesis on "Information aggregation and Political Economics". His main research interests are: opinion dynamics, evolutionary game theory, network theory and industrial organization. His e-mail address is: simone.righi@tk.mta.hu and up to date information and research papers can be found at his web-page: \url{http://perso.fundp.ac.be/~srighi/}.

{\bf K\'{A}ROLY TAK\'{A}CS} is Director of the "Lend\"ulet" Research Center for Educational and Network Studies (RECENS) of the Hungarian Academy of Sciences and associate professor of Sociology at the Corvinus University of Budapest. He obtained his Ph.D. in Sociology from the University of Groningen (Netherlands) with a thesis on "Social Networks and Intergroup Conflict". During and after his graduate studies he has been visiting scholar at Cornell University, University of Brescia and at the Netherlands Institute for Advanced Study in Humanities and Social Sciences. His main research interests are: social networks, intergroup conflict, discrimination, evolution of altruism and of cooperation, agent-based simulations and experiments. His e-mail address is: takacs.karoly@tk.mta.hu. Up to date information and research papers can be found at his web-page: \url{http://web.uni-corvinus.hu/~tkaroly/}.

\bibliographystyle{ws-acs}
\bibliography{RighiTakacsECMS}

\end{document}